\documentclass[prl,twocolumn,amsmath,showpacs,amssymb,floatfix]{revtex4}
\def\figwidth{0.9\linewidth}
\usepackage[final]{graphicx}

\include{psfig}
\input epsf

\begin{document}
\title{Activated bond-breaking processes preempt the observation of a sharp glass-glass transition in dense short-ranged attractive colloids}
\author{Emanuela Zaccarelli, Giuseppe Foffi, Francesco  Sciortino and Piero Tartaglia}
\address{ Dipartimento di Fisica, INFM UdR and Center for Statistical 
Mechanics and Complexity, Universit\`{a} di Roma La Sapienza, P.le
A. Moro 5, I-00185 Rome, Italy}
\begin{abstract}

We study --- using molecular dynamics simulations --- the temperature
dependence of the dynamics in a dense short-ranged attractive
colloidal glass to find evidence of the kinetic glass-glass transition
predicted by the ideal Mode Coupling Theory.  According to the theory,
the two distinct glasses are stabilized one by excluded volume and the
other by short-ranged attractive interactions. By studying the density
autocorrelation functions, we discover that the short-ranged
attractive glass is unstable.  Indeed, activated bond-breaking
processes slowly convert the attractive glass into the hard-sphere
one, preempting the observation of a sharp glass-glass transition.

\end{abstract}


\maketitle


Short-ranged attractive colloidal systems have recently become the
focus of many experimental\cite{poon-trappe} and
theoretical\cite{lekker} studies. The interest in these systems stems
from their peculiar dynamics\cite{sciortino02}, for showing structural
arrest phenomena both of gelation and vitrification type, and for
being amenable of analytic treatments. Previous studies have
convincingly shown that unusual dynamical phenomena emerge from the
competition between two characteristic localization length scales; the
hard-core and the short-ranged attraction localization lengths.  In
hard-sphere colloids, when the packing fraction $\phi$ exceeds roughly
$0.58$, the colloidal suspension turns into a glass and
particles are confined in cages of size of the order of 10$\%$ of the
hard core diameter (the hard-sphere localization length). In
attractive colloids, localization can be realized not only via
excluded volume (i.e. tuning $\phi$) but also thermally (i.e. tuning
the ratio between temperature $T$ and potential
depth)\cite{experiments}.  Hence, slowing down of the dynamics can be
induced not only by increasing $\phi$ but also by progressively
lowering $T$, i.e. increasing the inter-particle bonding.  Previous
studies have shown that when the range of the attractive interaction
is smaller than 10$\%$ of the hard-core diameter, an additional
localization length, controlled by the potential, sets in and an
efficient mechanism of competition between the hard-sphere and the
short-ranged bonding localization lengths arises.  In the fluid
phase-region where the two localization mechanisms compete, a highly
non trivial dynamics is observed: particle diffusion shows a maximum
on heating, the fluid can be transformed at constant $\phi$ into a glass
both on cooling and heating, the time dependence of dynamical
quantities like the mean squared displacement and the density-density
correlation function, have an uncommon behavior at intermediate times,
showing respectively a sub-diffusive and a logarithmic regime. These
anomalous properties have been recently observed in a series of
beautiful experiments \cite{mallamace,science02,bartsch02,malla2}, and
extensive simulations of particles interacting via short-ranged
square-well\cite{Rcomm02,zacca02,a4} and Asakura-Oosawa
potential\cite{puertas02}.

Hard sphere colloids have been an important system for accurately
testing theoretical predictions.  It has been shown that the ideal
Mode Coupling Theory (MCT), which neglects hopping phenomena, provides
an accurate description of the dynamics close to the hard-sphere glass
transition\cite{goetze91,vanmegen}.  Ideal MCT provides an accurate
description of the fluid phase dynamics also in short-ranged attractive
colloids, despite the complex dynamical processes alluded to previously.
Several theoretical predictions
\cite{fabbian99,bergenholtz99,dawson00,zaccarelli01,sperl02,sperlpisa}
have been recently quantitatively confirmed by
experiments\cite{mallamace,science02,bartsch02,malla2} and simulations
\cite{science02,Rcomm02,zacca02,puertas02,a4}.

An untested important prediction of the theory regards the presence of
a kinetic (as opposed to thermodynamics) glass-glass transition which
should take place in the glass phase on crossing a critical
temperature.  Heating a short-ranged attractive glass should produce a
sudden variation of all dynamical features, without significant
structural changes. For example, at the transition temperature, the
value of the long-time limit of the density-density correlation
function, the non-ergodicity factor $f_q$, should jump from the value
characteristic of the short-ranged attractive glass to the
significantly smaller value characteristic of the hard-sphere glass.

In this Letter we report an extensive numerical study of the binary
short-ranged square well (SW) potential in the {\it glass} phase
aiming at studying the $T$-evolution of the 
glass dynamics and detecting the 
location of the glass-glass transition. We discover that 
hopping phenomena destabilize the attractive glass
and prevent the observation of the 
a discontinuous glass-glass transition.

We perform molecular dynamics (MD) calculations of a $50\%-50\%$
binary mixture of $N=700$ particles of unitary mass interacting via a
square-well potential of unitary depth $-u_0$. The asymmetry between
the two hard-core diameters is fixed to $20\%$.  The attractive
well-width $\Delta_{ij}$ is given by $\Delta_{ij}/(\Delta_{ij}+\sigma_{ij}) = 0.03
$, where $\sigma_{ij}$ is the hard-core
diameter for the $ij$-type interaction\cite{units}, with $i,j=A,B$. 
In this model, a bond between two particles can be unambiguously defined when the  pair interaction energy is $-u_0$. The one-component
version of this model has been extensively studied within MCT
\cite{dawson00}.  The binary mixture case, which allows to prevent
crystallization without substantial changes in the dynamics, has been
studied numerically in the fluid phase \cite{zacca02}.  The
iso-diffusivity curves in the $\phi-T$ region are re-entrant and,
close to the re-entrance, correlation functions show a logarithmic
decay in agreement with MCT predictions.
Comparing the theoretical MCT calculations for this specific model, with the numerical results reported in Ref.\cite{zacca02} --- following the procedure first used by Sperl\cite{sperl03} --- the glass-glass transition line has been located
between  ($\phi \simeq 0.625$, $T \simeq 0.37$) and the   end-point ($\phi \simeq 0.64$, $T \simeq 0.42$)\cite{mapping}. At $\phi=0.635$, the packing fraction studied in this work, the glass-glass temperature is predicted to be $T \simeq 0.4$.

To generate glass configurations, we compress\cite{compress}, to
$\phi=0.635$, $136$ independent equilibrium configurations generated
at $T=0.6$ and $\phi\simeq 0.612$ in a previous study\cite{zacca02}.
We checked that the results presented in this Letter do not depend on
the $T$ and $\phi$ of the starting uncompressed configurations.  Since
the simulated system is out-of-equilibrium, ensemble averages are
requested to reduce the noise level.  Each compressed configuration
was instantly quenched to several temperatures $T$ by proper rescaling
of the initial velocities and let to evolve in a NVT
simulation\cite{thermostat}.  On the time scale of our study the
system behaves as a glass at all $T$ values, i.e. no diffusional
processes are observed.  We analyze nine different $T$ values, from
$T=1.5$ to $T=0.1$.  Each $T$ requested about 15 CPU days on fast
AMD Athlon processor. 

Since the system is in an out-of-equilibrium state, correlation
functions depend not only on $t$ but also on the time elapsed from the
compression, the waiting time $t_w$. For $t<<t_w$, correlation
functions for different $t_w$ collapse on the same curve --- a
phenomenon connected to the equilibration of the degrees of freedom
faster than $t_w$\cite{aging}. We exploit such feature to generate
$t_w$-independent data and focus only on the $T$-dependence. For this
reason we consider the dynamics for times up to approximately $400$
time units and $t_w > 4000$.

\begin{figure} 
 \includegraphics[width=\figwidth]{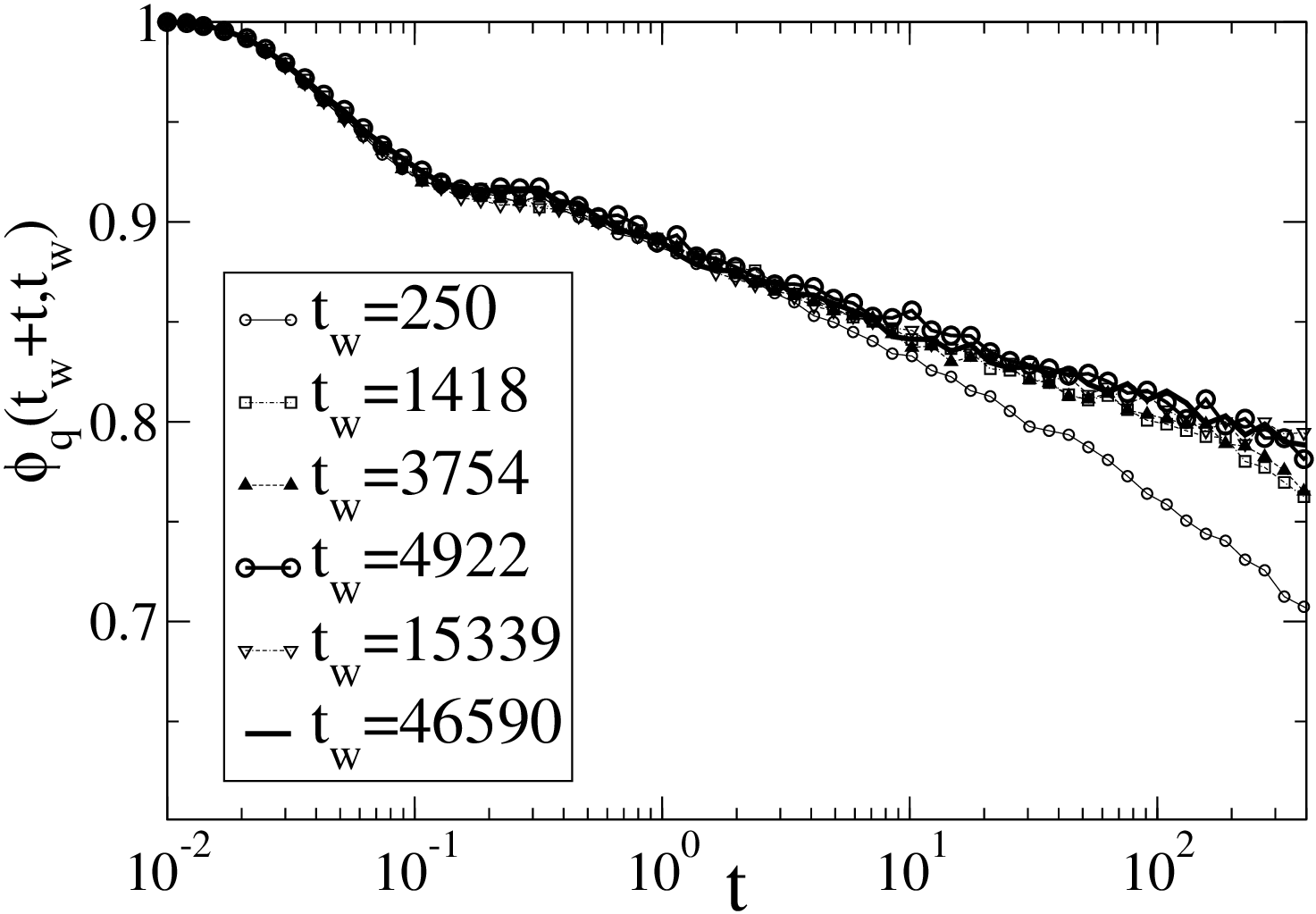}
\caption{ Waiting-time dependence of 
$\phi_q(t_w+t,t_w)$ for $T=0.35$, at $q\sigma_{BB}\simeq 25$. For $t_w > 4000$, no aging is observed in the chosen time window.   
Time is measured in reduced units \protect\cite{units}.}  
\label{fig:no-aging}  
\end{figure}

To confirm that  in the time window $0<t<400$, no $t_w$-dependence is observed for $t_w \gtrsim 4000$, we
calculate the density-density correlation function
\begin{equation}
\phi_{q}(t_w+t,t_w)\equiv \langle \rho^*_{\bf q}(t_w+t) \rho_{\bf q}(t_w) \rangle/\langle|\rho_{\bf q}(t_w)|^2\rangle
\end{equation}
where $\rho_{\bf q}(t)=\frac{1}{\sqrt{N}} \sum_i e^{i{\bf q} {\bf
r}_i(t)}$ and ${\bf r}_i(t)$ are the coordinates of particle $i$ at time
$t$.  Fig. \ref{fig:no-aging} shows $\phi_{q}(t_w+t,t_w)$ for the case
$T=0.35$, at $q\sigma_{BB}\simeq25$. For $t_w>4000$, $\phi_{q}(t_w+t,t_w)$ becomes independent of $t_w$ for $t \lesssim 400$. We confirmed such $t_w$-independence at all studied $T$-values.

\begin{figure} 
\includegraphics[width=\figwidth]{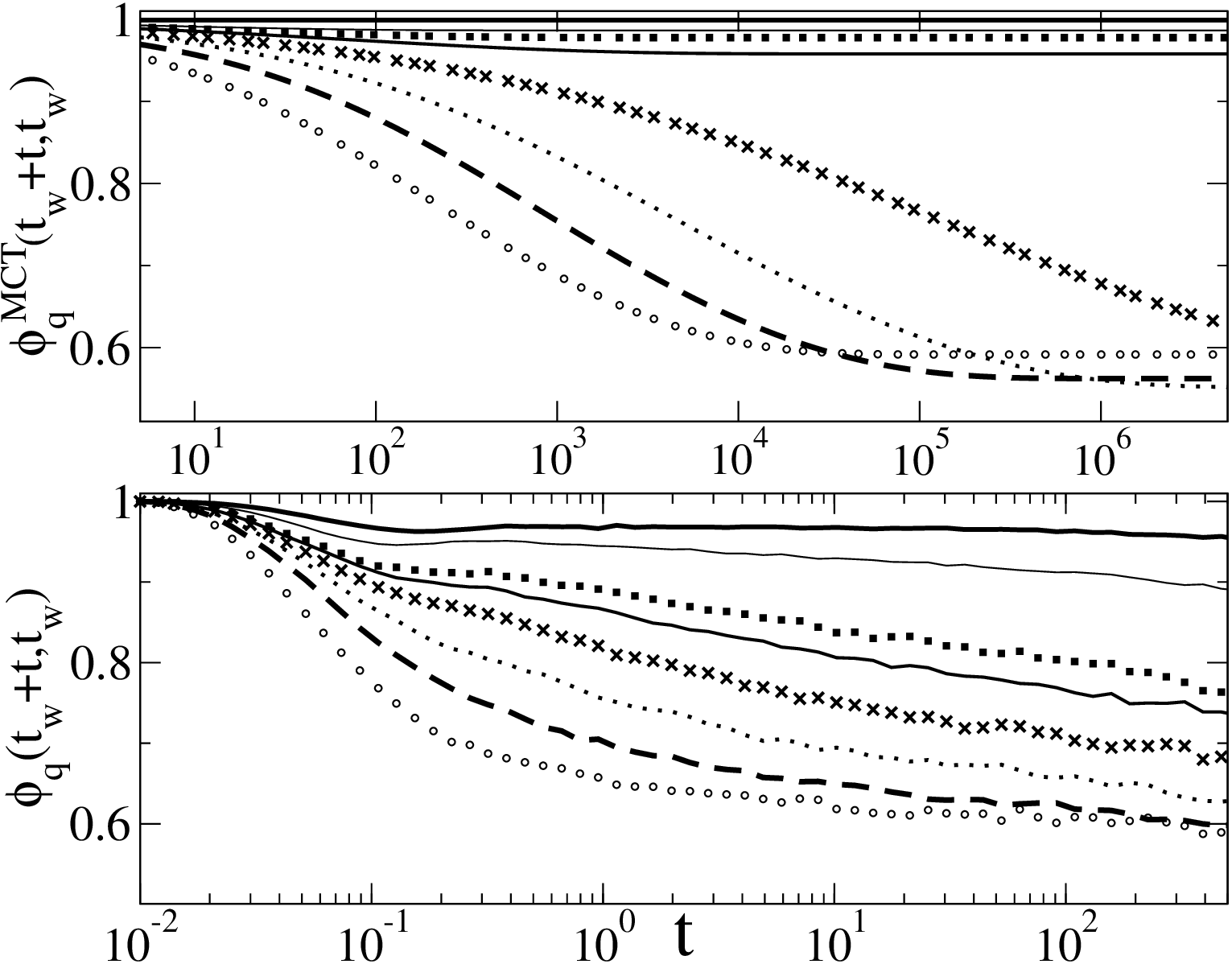}
\caption{
Top panel: Density correlators predicted by MCT, $\phi_q^{MCT}(t)$,
for a one-component SW system for different temperatures, crossing the
glass-glass transition at $\phi\simeq0.54$, for $q\sigma=14.5$. Such
$q$-value has been chosen to reproduce comparable differences in the
non-ergodicity parameters. The temperatures have been chosen such that
$T/T_c$ ($T_c$ is the MCT transition temperature) is close the
simulation one for the reported $T$'s. Bottom Panel: Same curves from
the simulation for $q\sigma_B\simeq25$. Temperatures from top to
bottom: $0.1,0.2,0.35,0.4,0.5,0.65,0.9,1.5$. $t_w=3754$. }
\label{fig:correlators}
\end{figure}

Before presenting the numerical data, we discuss the ideal-MCT
predictions for the one-component system, using the Percus-Yevick
static structure factor as input, extending the calculation reported
in Ref.\cite{dawson00}.  Fig. \ref{fig:correlators} (top panel) shows
$\phi_q^{MCT}(t)$ at different $T$ values. The long-time limit $f_q$
of the correlation function separates two groups of temperatures. At
high $T$ values, correlation functions tend to $f_q^{HS} \approx
0.6$, while for lower $T$, $f_q>0.9$, a figure typical of the
short-ranged attractive glass. The sharp jump in $f_q$ is the signature
of a discontinuous glass-glass transition.

Fig. \ref{fig:correlators} (bottom panel) shows $\phi_q(t_w+t,t_w)$ in
the $t_w$-independent time window, calculated from MD trajectories.
At the highest $T$, $T=1.5$, $\phi_q(t_w+t,t_w)$ quickly approaches a
plateau, consistent with the hard-sphere glass value, as expected
since the kinetic energy is larger than the depth of the square well.
On cooling, the approach to the plateau takes longer and longer,
consistently with the MCT predictions.  Around $T=0.4$, the decay is
well represented by a logarithmic law, in agreement with the location
of the MCT higher order singularity --- the source of the logarithmic
dynamics\cite{a4}.
For $T \lesssim 0.4 $, the correlation functions show a very
ill-defined plateau, around $0.9$, a value consistent with the MCT
prediction for the short-ranged attractive glass.  While in the
theoretical predictions the attractive plateau lasts for ever, in the simulated systems 
all correlation functions, independently from $T$, appear to decay well
below 0.9. The correlation functions for the two extreme temperatures in Fig.\ref{fig:correlators} (bottom panel) show that a
clear difference in the dynamics is indeed observed. 
However, the sharp glass-glass transition predicted by ideal MCT is not found.
These results   suggest that the
range of stability of the attractive glass is considerably shorter than
theoretically expected.  Even lowering $T$ down to $0.1$, i.e.  when
$T$ is one tenth of the potential well depth and the system is
expected to be deep in its attractive glass phase, the attractive
plateau lasts less than two decades in time\cite{considerazione}. 
We also note that the time window in which the correlation function is close
to the attractive plateau value does not extend  on  increasing $t_w$, as clearly shown  in Fig.\ref{fig:no-aging}. The absence of a clear short-ranged attractive plateau seems to suggest that the attractive cages are not as stable (in time) as the hard-sphere ones and that particles do manage to break the bond confinement\cite{shear}.

A possible explanation of the observed instability of the attractive glass can be formulated  in term of the so-called activated (or hopping) processes, decay processes which are not included in the ideal MCT\cite{extMCT}.  While in hard-spheres hopping phenomena are not significant due to the absence of an energy scale, in molecular systems it is  well known that the   "MCT glass" is  destabilized by activated processes, which slowly restore ergodicity. Indeed, in these systems the calorimetric glass transition temperature is located well below the MCT critical temperature.  In short-ranged attractive colloids, activated processes can be associated with thermal fluctuations of order $u_0$, which are able to break the bonds.  These processes generate a finite bond lifetime and destabilize the attractive glass. If $\phi$ is sufficiently large, the underlying hard-core repulsive glass should emerge as limiting disordered arrested structure.

\begin{figure} 
\includegraphics[width=\figwidth]{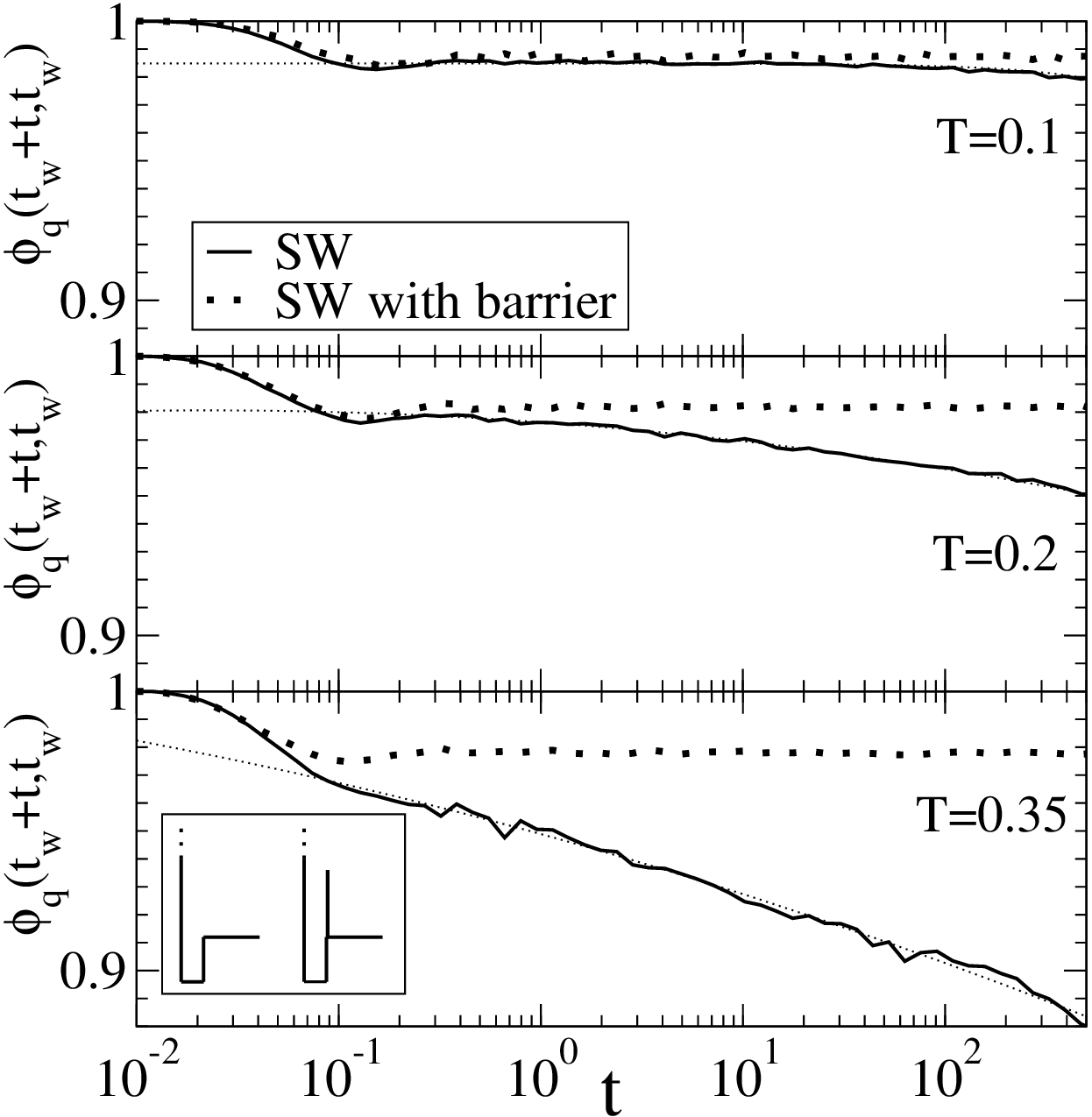}
\caption{ Comparison of the density correlators $\phi_q(t_w+t,t_w)$
for the square well (SW) model and  the SW model complemented by a  barrier of infinitesiman width (SWB) and  height 100$u_0$, as sketched in the figure.  $t_w=3754$.  The dashed lines are fits to the SW data, reported as a guide to the eye to help comparing the SW and SWB $f_q$ values.
}  \label{fig:barrier-dynamics}  
\end{figure}

It is tempting to formulate the hypothesis that an ideal MCT short-ranged attractive glass would be stable if bond-breaking processes were negligible.   Such ideal attractive glass should be  characterized by bond-cages with infinite lifetime, in analogy with the hard-sphere glass which is characterized by neighbor-cages of infinite lifetime. To substantiate this claim, we artificially build a model in which bonds are never broken on the time scale of our numerical observation.   To do so we add a finite barrier of infinitesimal width in the potential, just outside the attractive well (see Fig.\ref{fig:barrier-dynamics}), with the purpose to make the bonds longer lived, and solve the Newton equations exactly\cite{rapaport}. The nice feature of this modified system is that the thermodynamics is identical to the one of the original potential, since the width of the repulsive barrier is infinitesimal.  Dynamics are instead different, since bond-breaking
takes now place on longer time-scales; for example, by choosing as height of the
repulsive barrier 100$u_0$, we find that no bond is formed or broken in the
timescale of the simulation.  

Fig. \ref{fig:barrier-dynamics} contrasts the decay of
$\phi_q(t_w+t,t_w)$ for different $T$ values in the range of $T$ where
attractive glass is expected, both for the SW model and for the case
of the square well complemented by a barrier (SWB) of height
100$u_0$. The same initial configurations (at $t_w$) are used for both
models, so that at time $t=0$ the bonding pattern is identical. By
construction, only decorrelation processes which do not modify the
bonding pattern, i.e. the rattling motion within the bond-cages, are
possible in the SWB model.  We observe that, in the case of the SWB
model, the correlation functions decay to a stable plateau, whose
value, larger than 0.9, is consistent with the expected value for the
attractive glass.  In contrast, correlation functions for the
unmodified SW model, already at short times, decay below the plateau
value. Still, the extrapolated amplitude of the long-time decay (see
Fig. \ref{fig:barrier-dynamics}) provides an estimate of $f_q$ very
similar to the one of the SW model plus barrier. These results support
the view that ideal-MCT, by neglecting the bond-breaking processes,
predicts a stability window for the short-ranged attractive glass
larger than numerically observed.

The present study suggests  that the ideal attractive glass line in short-ranged
attractive colloids has to be considered, in full analogy with what
has been found in the study of glass-forming molecular liquids\cite{extMCT}, as a
cross-over line between a region where ideal-MCT predictions are
extremely good (in agreement with the previous calculation in the
fluid phase) and an activated-dynamics region, where ideal-MCT
predictions apply in a limited time window.  The anomalous dynamics
which stem from the presence of an higher-order singularity in the
MCT-equations\cite{fabbian99,a4}, still affect the dynamical processes in the fluid and
in the glass, even if activated processes pre-empt the possibility of
fully observing the glass-glass transition phenomenon, at least in the
SW case.  Short-ranged inter-particle potentials which stabilize bonding
could produce dynamics which are less affected by hopping processes,
favoring the observation of the glass-glass phenomenon\cite{malla2}.

We acknowledge support from MIUR COFIN 2002 and FIRB.
We thank S. Buldyrev for the MD code,  W. G\"otze for a critical reading of the manuscript and M. Fuchs and  A.M. Puertas for discussions. We acknowledge support  from INFM Iniziativa Calcolo Parallelo.



\begin{references}

\bibitem{poon-trappe} 
	P. N. Pusey, A. D. Pirie and W. C. K. Poon, Physica A {\bf 201},
	322 (1993); S. M. Ilett {\it et al}, Phys. Rev. E {\bf 51},
	1344 (1995); P. N. Segre {\it et al}, Phys. Rev. Lett. {\bf 86},
	6042 (2001); V. Trappe {\it et al}, Nature {\bf 411}, 772
	(2001).


\bibitem{lekker} 
	H. N. W. Lekkerkerker {\it et al}, Europhys. Lett. {\bf 20}, 559
	(1992); P. Bolhuis, M. Hagen abd D. Frenkel, Phys. Rev. E {\bf
	50}, 4880 (1994); C. F. Tejero {\it et al}, Phys. Rev. E {\bf
	51}, 558 (1995); V. J. Anderson and H. N. W. Lekkerkerker, Nature
	{\bf 416}, 811 (2002); K. A.  Dawson, Curr. Opin. Colloid Interf. {\bf 7}, 218 (2002).


\bibitem{sciortino02} 
F. Sciortino, Nature Materials {\bf 1}, 145 (2002).

\bibitem{experiments} In experiments, depletion interactions control the interparticle potential. The role of $T$ is provided by the amount of non-adsorbing polymer,
while the range of interaction is controlled by the length of the polymer.


\bibitem{mallamace}
        F. Mallamace {\it et al},  Phys. Rev. Lett. {\bf 84}, 5431 (2000).  



\bibitem{science02} 
        K. N. Pham {\it et al}, Science {\bf 296}, 104 (2002).


\bibitem{bartsch02} T. Eckert and E. Bartsch, Phys. Rev.
        Lett. {\bf 89}, 125701 (2002).

\bibitem{malla2} W. R. Chen, S. H. Chen and F. Mallamace, Phys. Rev. E
{\bf 66}, 021403 (2002).  S. H. Chen, W. R. Chen and F. Mallamace, Science, {\bf 300}, 619 (2003).



\bibitem{Rcomm02}G. Foffi {\it et al}, Phys. Rev. E {\bf 65}, 050802(R) (2002).



\bibitem{zacca02} E. Zaccarelli {\it et al}, Phys. Rev. E {\bf 66}, 041402 (2002).

\bibitem{a4}    F.  Sciortino, P.  Tartaglia, E.  Zaccarelli, cond-mat/0304192
(2003).




\bibitem{puertas02} 
        A. M. Puertas, M. Fuchs and M. E. Cates, Phys. Rev.
        Lett. {\bf 88}, 098301 (2002); Phys. Rev. E {\bf 67}, 031406 (2003).

\bibitem{goetze91} 
         W. G\"otze in 
         {\it Liquids, Freezing and Glass Transition} 
         edited by J.P. Hansen , D. Levesque D, and J. Zinn-Justin  
         (Amsterdan: North-Holland) p~287, 1991.

\bibitem{vanmegen}   
        W. van Megen and S. M. Underwood,  
        Phys. Rev. Lett. {\bf 70}, 2766 (1993);
        Phys. Rev. E {\bf 49}, 4206 (1994). 
 


\bibitem{fabbian99} L. Fabbian {\it et al}, Phys. Rev. E R1347 (1999), and
         Phys. Rev. E {\bf 60}, 2430 (1999).


\bibitem{bergenholtz99} 
         J. Bergenholtz and M. Fuchs, Phys. Rev. E {\bf 59}, 5706 
         (1999). 

\bibitem{dawson00}
        K. A. Dawson {\it et al}, Phys. Rev. E {\bf 63}, 011401 (2001).


\bibitem{zaccarelli01} 
          E. Zaccarelli {\it et al}, Phys. Rev. E {\bf 63}, 031501 (2001).

  


\bibitem{sperl02} 
	W. G\"{o}tze and M. Sperl, Phys. Rev. E {\bf 66},
	011405 (2002).

\bibitem{sperlpisa} 
	W. G\"{o}tze and M. Sperl, J. Phys. Condens. Matt. {\bf 15}, S869 (2003).


\bibitem{units} 
	Lengths are measured in units of $\sigma_B$, time in units of
	$ \sigma_B(m/u_0)^{1/2}$ and $k_B$, the Boltzmann constant, is
	fixed to $1$.  Energy and $T$ are thus measured in units of
	$u_0$.  The packing fraction $\phi$ is defined as $\pi/6 (N_A \sigma_A^3 + N_B \sigma_B^3)/V$, where $N_A=N_B=350$ and
	$V$ is the volume. 

\bibitem{sperl03}  
M. Sperl,
preprint (2003).

\bibitem{mapping} It is known that MCT overestimates
the tendency to form a glass. A mapping is required to compare theoretical and numerical results. In the present case, a bilinear mapping $(\phi\rightarrow 1.897\ \phi-0.3922, T\rightarrow 0.5882\ T-0.225)$ successfully superimposes the MCT predictions for the two ideal glass lines onto the corresponding numerical estimates, as discussed at length in Ref.\protect\cite{a4}.

\bibitem{compress} 
	To quickly compress, we run a simulation of a system of
	particles interacting with a hard-core followed by a repulsive
	shoulder. The configuration to be compressed is used as
	initial conditions in the Newton Equations.  The heigth of the
	shoulder is set to a value much larger than $T$, forcing
	closeby particles to quickly separate. As soon as the energy
	of the system goes back to zero, the simulation is interrupted
	and the configuration is saved. A proper rescaling of the
	coordinates produce a configuration consistent with the
	original square-well potential but with a larger packing
	fraction (fixed by the width of the repulsive shoulder). In
	this way, we were able to generate configurations up to about
	$\phi=0.65$.

\bibitem{thermostat} 
	It is worth observing that after a waiting time of quench of
	about $250$ MD units, no significant release/absorption of
	heat is observed in the studied time window.  This is due to
	the high packing fraction and the associated slowness of the
	particle dynamics which allows very minute configurational
	changes.


\bibitem{aging} 
	W. Kob and J.-L. Barrat, Phys. Rev. Lett. {\bf 78}, 4581 (1997);
	L. F. Cugliandolo, condmat/0210312.

\bibitem{considerazione} 
	The presence of a short-lifetime of the attractive glass cage
	was also encountered in \protect\cite{zacca02,puertas02}.

\bibitem{shear}
  Despite the fact that the lifetime of the
attractive glass is not infinite, the presence of a short-ranged
bonding significantly affects the time evolution of $\phi_q$ and,
correspondingly, the frequency dependence of the elastic modulus on
varying $T$.






\bibitem{extMCT} 
	S. P. Das and G. F. Mazenko, Phys. Rev. A {\bf 34}, 2265 (1986);
	W. G\"otze and L. Sj\"ogren, Z. Phys. B {\bf 65}, 415 (1987).





\bibitem{rapaport}         
We have implemented an event-driven algorithm, where in addition to hard-core collisions, the events include also the barrier crossing as explained in 
D. C. Rapaport, {\it The Art of Molecular Dynamic Simulation},    Cambridge University Press, 1995.



\end{references}
\end{document}